\documentclass[aps,prd,floats,floatfix,nofootinbib]{revtex4-1}
\usepackage{graphicx,amsmath}
\usepackage{amssymb}
\usepackage{amsfonts}

\begin{document}


\title{Yukawa alignment from natural flavor conservation}

\author{Graham~Cree}
\email{cree.graham@gmail.com}
\affiliation{Carleton University, Ottawa, Ontario K1S 5B6, Canada}

\author{Heather E.~Logan}
\email{logan@physics.carleton.ca}
\affiliation{Carleton University, Ottawa, Ontario K1S 5B6, Canada}

\date{August 25, 2011}

\begin{abstract}
We study the charged Higgs couplings to fermions in the ``democratic'' three-Higgs-doublet model, in which one doublet couples to down-type quarks, one to up-type quarks, and one to charged leptons.  Flavor-changing neutral Higgs couplings are absent because the Glashow-Weinberg-Paschos condition for natural flavor conservation is in effect.  We show that this model reproduces the coupling structure of the charged Higgs boson in the recently-proposed Yukawa-aligned two-Higgs-doublet model, with two subtle constraints that arise from the unitarity of the charged Higgs mixing matrix.  Adding a fourth Higgs doublet with no couplings to fermions eliminates these constraints.
\end{abstract}

\maketitle

\section{Introduction}

It has long been known that extending the Higgs sector of the Standard Model (SM) to include one or more additional Higgs doublets leads generically to flavor-changing neutral Higgs couplings, which are severely constrained by experiment.  The most commonly-applied way to suppress these flavor-changing couplings is to impose the condition of {\it natural flavor conservation}---proposed by Glashow, Weinberg and Paschos in 1977~\cite{GWP}---which stipulates that all right-handed fermions with a given electric charge couple to exactly one Higgs doublet.  With two Higgs doublets, this condition allows for four\footnote{We ignore neutrino masses.  Two-doublet models for Dirac neutrino masses have been constructed in Refs.~\cite{Fayet:1974fj,Davidson:2009ha} and \cite{Ma:2000cc,Wang:2006jy} with natural flavor conservation enforced by a U(1) or $Z_2$ symmetry, respectively.  The $Z_2$ model contains a very light scalar and has recently been shown to be strongly disfavored by astrophysical constraints~\cite{Zhou:2011rc}.  A supersymmetric version of the U(1) model was constructed in Ref.~\cite{Marshall:2009bk}.} different coupling assignments~\cite{Barnett:1983mm}: the usual Type I~\cite{Georgi:1978wr} and II~\cite{Lee:1973iz} two-Higgs-doublet models (2HDMs), as well as the less-well-known lepton-specific~\cite{Akeroyd:1996di,Barger:2009me,Aoki:2009ha,Goh:2009wg} and flipped~\cite{Akeroyd:1996di,Barger:2009me,Aoki:2009ha,Barger:1989fj,Logan:2010ag} 2HDMs (for a recent review, see Ref.~\cite{Branco:2011iw}).\footnote{The 2HDM without natural flavor conservation is called the Type III model~\cite{Cheng:1987rs}.  For a review of its phenomenology, see Refs.~\cite{Atwood:1996vj,Branco:2011iw}.}  Natural flavor conservation traces the absence of flavor-changing neutral Higgs couplings to the discrete symmetries that act on particular right-handed fermions and Higgs doublets in order to enforce the structure of the Yukawa Lagrangian.

An alternative approach to avoiding flavor constraints is to impose {\it minimal flavor violation}~\cite{MFV,Cirigliano:2005ck}.  Models with minimal flavor violation have the fermion flavor group (five copies of SU(3), for the three generations of each of $Q_L$, $u_R$, $d_R$, $L_L$, and $\ell_R$) broken only by the three usual Yukawa coupling matrices.\footnote{Some definitions of minimal flavor violation also require that the only CP violation in the model come from the phase in the SM CKM matrix; we do not impose that restriction.}  In a multi-Higgs-doublet model, therefore, minimal flavor violation attributes the absence of flavor-changing neutral Higgs couplings to the fundamental origin of the Yukawa matrices themselves.  The weak-scale Higgs doublets do not carry flavor-distinguishing quantum numbers.\footnote{We consider only color-singlet scalar doublets; MFV Yukawa couplings are also allowed for color octet scalars~\cite{Manohar:2006ga}.}

Because of the very different light that these two possibilities could shed on the origin of flavor violation, it is interesting to consider the prospects for distinguishing them experimentally.
The simplest implementation of minimal flavor violation in an extended Higgs sector, the so-called Yukawa-aligned 2HDM, was introduced in Ref.~\cite{Pich:2009sp}.  In this model, both Higgs doublets couple to all types of fermions; flavor-changing neutral Higgs interactions are avoided by requiring that the two Yukawa matrices that couple right-handed fermions of a given electric charge to the two Higgs doublets are proportional to each other,\footnote{Minimal flavor violation also admits a nonlinear realization~\cite{Kagan:2009bn}, in which couplings involve higher powers or products of Yukawa matrices.  This leads to nontrivial effects in the third generation due to the large top quark Yukawa coupling.  Here we consider only the linear Yukawa-aligned implementation.}
so that they are both diagonal in the fermion mass basis.\footnote{Steps toward an explicit implementation using family symmetries were taken recently in Ref.~\cite{Varzielas:2011jr}.}
The key free parameters in the model are the three complex proportionality constants between the three pairs of Yukawa matrices.  These parameters, comprising three magnitudes and two physically-meaningful phases, generalize the role of the usual $\tan\beta$ parameter in 2HDMs with natural flavor conservation.  As a laboratory for their effects, we focus on the couplings of the charged Higgs boson $H^+$ of the Yukawa-aligned 2HDM.  These couplings, and their experimental constraints, have been studied in detail in Ref.~\cite{Jung:2010ik}.

In this paper we show that the charged Higgs coupling structure of the Yukawa-aligned 2HDM can be mimicked in a three-Higgs-doublet model (3HDM) with natural flavor conservation.\footnote{In Ref.~\cite{Serodio:2011hg} it was recently shown that a constrained version of the Yukawa-aligned structure, with two of the three coupling parameters equal to each other and no complex phases, could be achieved in a multi-Higgs-doublet model with natural flavor conservation in which only two of the doublets couple to fermions.}  The study of the structure of charged Higgs couplings in a model with more than two Higgs doublets was pioneered by Albright, Smith and Tye in Ref.~\cite{Albright:1979yc} and by Grossman in Ref.~\cite{Grossman:1994jb}.  We will adopt the notation of Ref.~\cite{Grossman:1994jb}.
In order to obtain three independently-varying charged Higgs coupling magnitudes as in the Yukawa-aligned 2HDM, we couple the first of our three Higgs doublets to down-type quarks, the second to up-type quarks, and the third to charged leptons---the so-called ``democratic'' 3HDM.  
Charged Higgs couplings in this model, and the resulting constraints and experimental signatures, were first studied in Refs.~\cite{Grossman:1994jb,Akeroyd:1994ga}.
Neutral Higgs couplings in the CP-conserving version of this model were studied in Ref.~\cite{Barger:2009me}.  A supersymmetric version of this model, containing also a fourth Higgs doublet with no couplings to fermions, was introduced in Ref.~\cite{Marshall:2010qi}.  We show that the couplings of either of the two charged Higgs bosons in the (non-supersymmetric) 3HDM can be made to reproduce the charged Higgs couplings in the Yukawa-aligned 2HDM, with two subtle constraints that arise from the unitarity of the charged Higgs mixing matrix: first, that the three coupling parameter magnitudes cannot all be enhanced or all be suppressed simultaneously; and second, that the second complex phase of the Yukawa-aligned 2HDM is not a free parameter in the democratic 3HDM but instead is fixed in terms of the other three magnitudes and one phase.  We also show that adding a fourth Higgs doublet, with no couplings to fermions, eliminates these constraints by enlarging the charged Higgs mixing matrix.

The rest of this paper is organized as follows. In Sec.~\ref{sec:a2hdm} we briefly review the Yukawa-aligned 2HDM.  In Sec.~\ref{sec:d3hdm} we lay out the democratic 3HDM and define our notation, following Grossman~\cite{Grossman:1994jb}.  In Sec.~\ref{sec:unitarity} we derive our main results, which are the constraints on the charged Higgs coupling parameters in the democratic 3HDM.  In Sec.~\ref{sec:expt} we summarize the experimental constraints on the charged Higgs coupling parameters from previous literature.  In Sec.~\ref{sec:H3} we derive the couplings of the other charged Higgs in the democratic 3HDM and show that they are predicted entirely in terms of measurable couplings of the first charged Higgs.  In Sec.~\ref{sec:4HDM} we discuss the consequences of adding a fourth doublet with no couplings to fermions, and show that it removes the constraints on the charged Higgs coupling parameters.  In Sec.~\ref{sec:conclusions} we summarize our conclusions and briefly discuss other approaches to detecting the presence of a third Higgs doublet using the couplings of the neutral Higgs bosons.  The scalar potential and resulting charged scalar mass matrix for the democratic 3HDM are given in an appendix.


\section{Yukawa-aligned two-Higgs-doublet model}
\label{sec:a2hdm}

The Yukawa-aligned 2HDM~\cite{Pich:2009sp} contains two scalar SU(2)$_L$ doublets $\phi_i$, with generically complex vacuum expectation values (vevs) $v_i$.  Each doublet couples to all types of fermions via the Yukawa Lagrangian,
\begin{equation}
	\mathcal{L} = -\left\{ \bar{Q}_L (\Gamma_1 \phi_1 + \Gamma_2 \phi_2) d_R 
	+ \bar{Q}_L ( \Delta_1 \tilde{\phi}_1 + \Delta_2 \tilde{\phi}_2) u_R 
	+ \bar{L}_L (\Pi_1 \phi_1 + \Pi_2 \phi_2) \ell_R + {\rm h.c.} \right\},
\label{firstLagrangian}
\end{equation}
where $\Gamma_i$, $\Delta_i$, and $\Pi_i$ are $3\times 3$ complex Yukawa matrices and $\tilde \phi_i \equiv i \sigma_2 \phi_i^*$ is the conjugate doublet of $\phi_i$.  Tree-level flavor-changing neutral Higgs interactions are eliminated by imposing the linear version of minimal flavor violation called Yukawa alignment; i.e., by requiring that $\Gamma_1 \propto \Gamma_2$, $\Delta_1 \propto \Delta_2$, and $\Pi_1 \propto \Pi_2$.

Rotating to the Higgs basis, in which one doublet $H_1$ carries a nonzero real vev $v_{\rm SM} = \sqrt{|v_1|^2 + |v_2|^2} \simeq 246$~GeV and the other doublet $H_2$ has no vev, the Yukawa Lagrangian becomes
\begin{equation}
	\mathcal{L} = - \frac{\sqrt{2}}{v_{\rm SM}} 
	\left\{ \bar{Q}_L (\mathcal{M}_d H_1 + \mathcal{Y}_d H_2) d_R 
	+ \bar{Q}_L ( \mathcal{M}_u \tilde{H}_1 + \mathcal{Y}_u \tilde{H}_2) u_R 
	+ \bar{L}_L ( \mathcal{M}_{\ell} H_1 + \mathcal{Y}_{\ell} H_2) \ell_R + {\rm h.c.} \right\}.
\label{secondLagrangian}
\end{equation}
Here $\mathcal{M}_f$ are the undiagonalized mass matrices for fermions of type $f$ and $\mathcal{Y}_f/v_{\rm SM}$ are Yukawa matrices coupling $H_2$ to fermions.  Yukawa alignment forces $\mathcal{Y}_f \propto \mathcal{M}_f$, so that the $\mathcal{Y}_f$ matrices are automatically diagonalized in the fermion mass basis.  In particular, following Ref.~\cite{Pich:2009sp} we define
\begin{equation}
	\mathcal{Y}_f = \zeta_f \mathcal{M}_f.
\end{equation}
The three complex parameters $\zeta_f$ characterize the model and control the charged Higgs couplings.  There are five real free parameters, the magnutudes of the three $\zeta_f$ and two phases---one overall phase can be absorbed by a rephasing of $H_2$, and is thus not physically meaningful.  Using this freedom, we will choose $\zeta_{\ell}$ to be real.

In the Higgs basis, the charged Higgs boson lives entirely in $H_2$.  Its couplings to fermions are thus controlled by $\mathcal{Y}_f$.  In the fermion mass basis, the charged Higgs Yukawa Lagrangian is given in terms of the diagonalized fermion mass matrices $M_f$, the CKM matrix $V$, and the parameters $\zeta_f$ by,\footnote{Here the minus sign in front of $\zeta_u$ comes from the extra minus sign on the charged scalar in the conjugate doublet $\tilde H_2$.  We define the neutrinos in the flavor eigenbasis.}
\begin{equation}
	\mathcal{L} = - \frac{\sqrt{2}}{v_{\rm SM}} 
	\left[ \zeta_d \bar{u}_L V M_d d_R - \zeta_u \bar{u}_R M_u V d_L 
	+ \zeta_\ell \bar{\nu}_L M_\ell \ell_R \right] H^+ + {\rm h.c.}
\label{eq:zetalag}
\end{equation}


\section{Democratic three-Higgs-doublet model}
\label{sec:d3hdm}

The democratic 3HDM employs natural flavor conservation to eliminate tree-level flavor-changing neutral Higgs couplings.  The model contains three scalar SU(2)$_L$ doublets, denoted $\Phi_d$, $\Phi_u$, and $\Phi_{\ell}$, with
\begin{equation}
	\Phi_f = \left( \begin{array}{c} \phi_f^+ \\ (v_f + \phi_f^{0,r} + i \phi_f^{0,i})/\sqrt{2} 
	\end{array} \right).
\end{equation}
The vevs $v_f$ of the three Higgs doublets can be chosen real through an independent rephasing of each doublet.  They are constrained by the $W$ boson mass to satisfy $v_{\rm SM} = \sqrt{v_d^2 + v_u^2 + v_{\ell}^2} \simeq 246$~GeV.

In order to enforce natural flavor conservation, we introduce three $Z_2$ symmetries, under which the charges of the Higgs doublets and SM fermions are given in Table~\ref{tab:Z2charges}.   
\begin{table}
\begin{tabular}{cccc}
\hline\hline
Field & $Z_2^d$ & $Z_2^u$ & $Z_2^{\ell}$ \\
\hline
$\Phi_d$ & $-$ & $+$ & $+$ \\
$d_R$ & $-$ & $+$ & $+$ \\
$\Phi_u$ & $+$ & $-$ & $+$ \\
$u_R$ & $+$ & $-$ & $+$ \\
$\Phi_{\ell}$ & $+$ & $+$ & $-$ \\
$\ell_R$ & $+$ & $+$ & $-$ \\
$Q_L$, $L_L$ & $+$ & $+$ & $+$ \\
\hline\hline
\end{tabular}
\caption{Charges of the three Higgs doublets and SM fermions under the three $Z_2$ symmetries imposed to enforce natural flavor conservation in the democratic 3HDM.}
\label{tab:Z2charges}
\end{table}
This choice forces $\Phi_d$ to couple only to $d_R$, $\Phi_u$ to $u_R$, and $\Phi_{\ell}$ to $\ell_R$.  The Yukawa Lagrangian takes the form,
\begin{equation}
	\mathcal{L} = - \left\{ \bar{Q}_L \Phi_d \mathcal{G}_d d_R
	+ \bar{Q}_L \tilde{\Phi}_u \mathcal{G}_u u_R 
	+   \bar{L}_L \Phi_\ell \mathcal{G}_\ell \ell_R + {\rm h.c.} \right\}.
\label{eq:ggg}
\end{equation}
Here $\mathcal{G}_f$ are $3\times 3$ complex Yukawa matrices related to the fermion mass matrices by $\mathcal{M}_f = \mathcal{G}_f v_f/\sqrt{2}$.  In the fermion mass basis, then, the Yukawa couplings are determined in terms of the corresponding fermion mass and the relevant vev $v_f$.  As in the usual 2HDMs with natural flavor conservation, we will assume that the $Z_2$ symmetries are softly broken (by dimension-two terms) in the Higgs potential; details of the potential are given in the appendix.

The final ingredient needed to determine the charged Higgs couplings is to specify the charged Higgs mass eigenstates.  The three charged fields $\phi_d^+$, $\phi_u^+$, and $\phi_{\ell}^+$ mix to form one charged Goldstone boson $G^+$ and two physical charged Higgs states, which we denote $H_2^+$ and $H_3^+$ following Grossman~\cite{Grossman:1994jb}, via a unitary mixing matrix $U$:\footnote{Complex phases in $U$ arise from CP-violating phases in the Higgs potential.}
\begin{equation}
	\left( \begin{array}{c} G^+ \\ H_2^+ \\ H_3^+ \end{array} \right) 
	= U \left( \begin{array}{c} \phi_d^+ \\ \phi_u^+ \\ \phi_\ell^+ \end{array} \right).
	\label{eq:Udef}
\end{equation}
All the information needed to determine the charged Higgs couplings is encoded in $U$.  First, the first row of $U$ is fixed by the composition of the Goldstone boson:\footnote{The charged Goldstone boson is uniquely determined as the linear combination of the fields $\phi_f^+$ that participates in the (unphysical) $W_{\mu}^- \partial^{\mu} G^+$ interaction coming from the scalar gauge-kinetic terms.  In any extended Higgs sector, the composition of $G^+$ therefore depends only on the vevs of the Higgs fields and the appropriate gauge generators.}
\begin{equation}
	G^+ = \left( v_d \phi_d^+ + v_u \phi_u^+ + v_{\ell} \phi_{\ell}^+ \right)/v_{\rm SM},
\end{equation}
so that $U_{1f} = v_f/v_{\rm SM}$, $f = d, u, \ell$.  We have already used the phase freedom of $\Phi_f$ to choose $v_f$, and hence all three $U_{1f}$, to be real.  

The couplings of the charged Higgs state $H_i^+$ are controlled by the relevant Yukawa coupling $\mathcal{G}_f$ and the overlap of the charged Higgs state with the relevant $\Phi_f$.  They can thus be written using Eqs.~(\ref{eq:ggg}) and (\ref{eq:Udef}) in terms of the fermion mass and elements of $U$.  Again following Grossman~\cite{Grossman:1994jb}, we define parameters $X_i$, $Y_i$, and $Z_i$ as follows,\footnote{The minus sign in the definition of $Y_i$ is included to simplify the Lagrangian in Eq.~(\ref{eq:xyzlag}) by taking into account the extra minus sign on the charged scalar in the conjugate doublet $\tilde \Phi_u$.}
\begin{equation}
	X_i = \frac{U_{di}^\dagger}{U_{d1}^\dagger}, \quad \quad 
	Y_i = - \frac{U_{ui}^\dagger}{U_{u1}^\dagger}, \quad \quad 
	Z_i = \frac{U_{\ell i}^\dagger}{U_{\ell 1}^\dagger},
\label{eq:xyz}
\end{equation}
with $i = 2,3$ corresponding to charged Higgs states $H^+_{2,3}$.  
With this notation the charged Higgs couplings become,
\begin{eqnarray}
	\mathcal{L} = - \frac{\sqrt{2}}{v_{\rm SM}} && \left\{ 
	\left[ X_2 \bar{u}_L V M_d d_R + Y_2 \bar{u}_R M_u V d_L 
	+ Z_2 \bar{\nu}_L M_{\ell} \ell_R \right] H_2^+ \right. \nonumber \\
	&& \left. + \left[ X_3 \bar{u}_L V M_d d_R + Y_3 \bar{u}_R M_u V d_L 
	+ Z_3 \bar{\nu}_L M_{\ell} \ell_R \right] H_3^+ + {\rm h.c.} \right\}.
\label{eq:xyzlag}
\end{eqnarray}

Let us first assume that $H_2^+$ is relatively light while $H_3^+$ is much heavier.  Then we can consider the couplings of $H_2^+$ in isolation.  Comparing the couplings of $H_2^+$ in Eq.~(\ref{eq:xyzlag}) to Eq.~(\ref{eq:zetalag}) for the charged Higgs couplings in the Yukawa-aligned 2HDM, we see that we can identify
\begin{equation}
	X_2 = \zeta_d, \qquad Y_2 = - \zeta_u, \qquad Z_2 = \zeta_\ell.
\end{equation}

Once again, $X_2$, $Y_2$, and $Z_2$ are three different complex parameters, one of which can be chosen real by rephasing $H_2^+$ (we choose $Z_2$ to be real).  It would appear that the phenomenology of $H_2^+$ in the democratic 3HDM is exactly the same as that of the charged Higgs in the Yukawa-aligned 2HDM!  But do the parameters $X_2$, $Y_2$, and $Z_2$ have the same freedom as the $\zeta_f$ in the Yukawa-aligned 2HDM?  The answer is {\it no}, as can be seen immediately by recalling that, after absorbing five phases into the definitions of $\phi_f^+$ and $H_i^+$, the $3 \times 3$ unitary matrix $U$ depends on only four real parameters (three angles and a phase). Starting from the basis in which the vevs $v_f$ are all real, we can define 
\begin{equation}
	\tan\beta = v_u/v_d, \qquad \tan\gamma = \sqrt{v_d^2 + v_u^2}/v_{\ell}.
\end{equation}
Then the matrix $U$ can be written explicitly as:\footnote{A similar parameterization was given in Ref.~\cite{Albright:1979yc}.}
\begin{eqnarray}
	U &=& \left( \begin{array}{ccc} 
		1 & 0 & 0 \\
		0 & e^{-i \delta} & 0 \\
		0 & 0 & 1 \end{array} \right)
		\left( \begin{array}{ccc}
		1 & 0 & 0 \\
		0 & c_\theta & s_\theta e^{i \delta} \\
		0 & -s_\theta e^{-i \delta} & c_\theta \end{array} \right)
		\left( \begin{array}{ccc}
		s_\gamma & 0 & c_\gamma \\
		0 & 1 & 0 \\
		-c_\gamma & 0 & s_\gamma \end{array} \right)
		\left( \begin{array}{ccc}
		c_\beta & s_\beta & 0 \\
		-s_\beta & c_\beta & 0 \\
		0 & 0 & 1 \end{array} \right)
	\nonumber \\
	&=& \left( \begin{array}{ccc}
	s_\gamma c_\beta & s_\gamma s_\beta 	& c_\gamma \\
	-c_\theta s_\beta e^{-i\delta} - s_\theta c_\gamma c_\beta 
		& c_\theta c_\beta e^{-i\delta} - s_\theta c_\gamma s_\beta & s_\theta s_\gamma \\
	s_\theta s_\beta e^{-i\delta} - c_\theta c_\gamma c_\beta 
		& -s_\theta c_\beta e^{-i\delta} - c_\theta c_\gamma s_\beta & c_\theta s_\gamma 
	\end{array} \right).
	\label{eq:Uexplicit}
\end{eqnarray}
Here $s$, $c$ denote the sine or cosine of the respective angle, $\theta$ is an angle describing mixing between $H_2^+$ and $H_3^+$, and $\delta$ is the CP-violating phase.  From right to left in the first line, the first two matrices fix the Goldstone boson $G^+$, while the third accomplishes the diagonalization of the (generally complex) mass matrix for the two physical charged Higgs bosons.  The matrix on the far left contains the phase rotation of $H_2^+$ that makes $Z_2$ real.  We have also chosen the phase of $H_3^+$ to make $Z_3$ real.

Thus the second free phase of the Yukawa-aligned 2HDM will be fixed in terms of the other four real parameters in the democratic 3HDM.  The explicit form of this constraint, together with another less-obvious constraint on the magnitudes of $X_2$, $Y_2$, and $Z_2$, are most easily seen by taking advantage of the unitarity of the matrix $U$.

\section{Unitarity constraints on the charged Higgs couplings}
\label{sec:unitarity}

Unitarity of the charged Higgs mixing matrix requires that
\begin{equation}
	\sum_f U_{if} U_{fj}^\dagger = \delta_{ij}.
\label{eq:unitar}
\end{equation}
Setting $i = j = 1$, we recover the sum rule for the vevs of the three doublets:
\begin{equation}
	|U_{1d}|^2 + |U_{1u}|^2 + |U_{1 \ell}|^2 
	= \frac{v_d^2}{v_{\rm SM}^2} + \frac{v_u^2}{v_{\rm SM}^2} + \frac{v_{\ell}^2}{v_{\rm SM}^2}
	= 1.
\label{eq:11}
\end{equation}
Setting $i = j = 2$ and using the definitions of $X_2$, $Y_2$, and $Z_2$ from Eq.~(\ref{eq:xyz}), we obtain a nontrivial constraint on the magnitudes of $X_2$, $Y_2$, and $Z_2$:
\begin{equation}
	|X_2|^2 |U_{1d}|^2 + |Y_2|^2 |U_{1u}|^2 + |Z_2|^2 |U_{1 \ell}|^2 = 1.
\label{eq:22}
\end{equation}
For a given choice of vevs, Eq.~(\ref{eq:22}) defines a plane in the three-dimensional positive-definite parameter space of $|X_2|^2$, $|Y_2|^2$, $|Z_2|^2$, passing through the point $(1,1,1)$, and intersecting the $|X_2|^2$ axis at $|U_{1d}|^{-2}$, the $|Y_2|^2$ axis at $|U_{1u}|^{-2}$, and the $|Z_2|^2$ axis at $|U_{1 \ell}|^{-2}$ (all three of these intersection point values are greater than 1).  This yields an interesting constraint on the parameters $X_2$, $Y_2$, and $Z_2$: 
\begin{quote}
The magnitudes of the coupling strengths $X_2$, $Y_2$, and $Z_2$ may not all be simultaneously less than one or simultaneously greater than one.
\end{quote}

We now derive the relationship that fixes the phase of $Y_2$ in terms of the other four parameters.  Setting $i = 1$ and $j = 2$ in Eq.~(\ref{eq:unitar}) and using the definitions of $X_2$, $Y_2$, and $Z_2$ from Eq.~(\ref{eq:xyz}), we obtain,
\begin{equation}
	X_2 |U_{1d}|^2 - Y_2 |U_{1u}|^2 + Z_2 |U_{1\ell}|^2 = 0.
	\label{eq:12}
\end{equation}
Because $X_2$ and $Y_2$ are complex, this represents two constraints: the real part and the imaginary part of the left-hand side must be separately equal to zero.  Together, Eqs.~(\ref{eq:11}), (\ref{eq:22}), and (\ref{eq:12}) constitute four constraints on eight real parameters ($U_{1d}$, $U_{1u}$, $U_{1\ell}$, $|X_2|$, $|Y_2|$, $Z_2$, and the phases of $X_2$ and $Y_2$), leaving four independent real free parameters (the usual three angles and a phase).

We begin by solving for the normalized vevs $|U_{1f}|^2 = v_f^2/v_{\rm SM}^2$.  Trivially from Eq.~(\ref{eq:11}) we have,
\begin{equation}
	v_{\ell}^2/v_{\rm SM}^2 = |U_{1\ell}|^2 = 1 - |U_{1d}|^2 - |U_{1u}|^2.
\end{equation}
Using this, we can solve both Eqs.~(\ref{eq:22}) and (\ref{eq:12}) for $|U_{1u}|^2$:
\begin{eqnarray}
	v_u^2/v_{\rm SM}^2 = |U_{1u}|^2
	&=& \frac{1 - |Z_2|^2 - |U_{1d}|^2 (|X_2|^2 - |Z_2|^2)}{|Y_2|^2 - |Z_2|^2},
	\label{eq:Uu1b} \\
	v_u^2/v_{\rm SM}^2 = |U_{1u}|^2 
	&=& \frac{Z_2 + |U_{1d}|^2 (X_2 - Z_2)}{Y_2 + Z_2}.
	\label{eq:Uu1a}
\end{eqnarray}
Equating the right-hand sides of Eqs.~(\ref{eq:Uu1b}) and (\ref{eq:Uu1a}) presents a solution for $|U_{1d}|^2$.  However, it does more than that.  The right-hand side of Eq.~(\ref{eq:Uu1a}) is complex, which means we have two real equations to solve and hence two independent solutions for $|U_{1d}|^2$ which must be simultaneously true.  These solutions are,
\begin{eqnarray}
	v_d^2/v_{\rm SM}^2 = |U_{1d}|^2 
	&=& \frac{{\rm Re} Y_2 (1 - |Z_2|^2) + Z_2 (1 - |Y_2|^2)}
	{{\rm Re} X_2 (|Y_2|^2 - |Z_2|^2) + {\rm Re} Y_2 (|X_2|^2 - |Z_2|^2) + Z_2 (|X_2|^2 - |Y_2|^2)},
	\label{eq:Ud1a} \\ 
	v_d^2/v_{\rm SM}^2 = |U_{1d}|^2 
	&=& \frac{{\rm Im} Y_2 (1 - |Z_2|^2)}
	{{\rm Im} X_2 (|Y_2|^2 - |Z_2|^2) + {\rm Im} Y_2 (|X_2|^2 - |Z_2|^2)},
	\label{eq:Ud1b}
\end{eqnarray}
where we have used the fact that $Z_2$ is chosen real to simplify the expressions.  Setting the right-hand sides of Eqs.~(\ref{eq:Ud1a}) and (\ref{eq:Ud1b}) equal, one can solve, e.g., for ${\rm Im} Y_2$ in terms of the other four real parameters.

\section{Experimental constraints}
\label{sec:expt}

We now briefly summarize the experimental constraints on the couplings of $H_2^+$ from existing studies.  Aside from the charged Higgs direct-search constraints, these arise from virtual exchange of the charged Higgs at tree or one-loop level.  Obtaining constraints on $X_2$, $Y_2$, and $Z_2$ requires the assumption that $H_3^+$ exchange does not contribute significantly to these processes; we thus continue to assume that $H_3^+$ is much heavier than $H_2^+$.  Determination of the combined constraints when the $H_2^+$ and $H_3^+$ masses are comparable would require a dedicated analysis, which is beyond the scope of this paper.

The experimental constraints on the charged Higgs couplings in the Yukawa-aligned 2HDM were comprehensively studied in Ref.~\cite{Jung:2010ik}.  Because of the correspondence between the Yukawa-aligned model and the democratic 3HDM, these constraints apply equally well to the latter (in the limit that effects due to $H_3^+$ can be neglected).  All constraints quoted are 95\% confidence level exclusions.

The strongest constraint is on $|Y_2|$ and comes from the LEP measurement of $R_b$ (the $b \bar b$ fraction in hadronic $Z$ boson decays); assuming that $|X_2| < 50$ so that contributions involving the bottom Yukawa coupling are not important, Ref.~\cite{Jung:2010ik} finds,
\begin{equation}
	|Y_2| \leq 0.72 + 0.24 \left( \frac{M_{H_2^+}}{100~{\rm GeV}} \right).
	\label{eq:Y2constraint}
\end{equation}
We note that if $M_{H_2^+} \simeq 100$~GeV, $|Y_2|$ must be less than one.  The unitarity constraint on the magnitudes of the couplings in the democratic 3HDM then dictates that at least one of $X_2$, $Z_2$ must be greater than one.

Lepton flavor universality in $\tau$ decays to $\mu$ versus $e$ provides the strongest constraint on $Z_2$; Ref.~\cite{Jung:2010ik} finds
\begin{equation}
	Z_2 \leq 40 \left( \frac{M_{H_2^+}}{100~{\rm GeV}} \right).
	\label{eq:Z2constraint}
\end{equation}

Constraints on products of couplings come from leptonic decays of heavy mesons.  $B \to \tau \nu$ yields an allowed annulus in the complex plane of $X_2 Z_2$, with an absolute upper bound of~\cite{Jung:2010ik}
\begin{equation}
	\left| X_2 Z_2 \right| \leq 1080 \left( \frac{M_{H_2^+}}{100~{\rm GeV}} \right)^2.
\end{equation}
Combining the constraints on $|Y_2|$ and $Z_2$ in Eqs.~(\ref{eq:Y2constraint}) and (\ref{eq:Z2constraint}), together with the LEP lower bound of about 79.3~GeV~\cite{Heister:2002ev} on the charged Higgs mass\footnote{This charged Higgs mass bound assumes that the charged Higgs decays only to a combination of $c \bar s$, $c \bar b$, and $\tau \nu$.  The strongest overall bound comes from ALEPH~\cite{Heister:2002ev}.  In the case that the branching fraction to $\tau \nu$ is close to 1, a stronger bound of 92.0~GeV comes from OPAL~\cite{Abbiendi:2003ji}.}, yields a constraint on the product $|Y_2 Z_2|$ which is much stronger than the semileptonic meson decay constraints~\cite{Jung:2010ik}.

The radiative decay $\bar B \to X_s \gamma$ receives charged-Higgs contributions with terms in the amplitude proportional to $|Y_2|^2$ and to $X_2 Y_2^*$.  Detailed combined constraints on $|Y_2|^2$ and the real and imaginary parts of $X_2 Y_2^*$ are presented in Ref.~\cite{Jung:2010ik}.  If $|Y_2|$ is not too big, as favored by the constraint from $R_b$, the 1$\sigma$ constraint on the real part of $X_2 Y_2^*$ was also given in a convenient form in Ref.~\cite{Trott:2010iz} neglecting contributions from ${\rm Im}(X_2Y_2^*)$.  The constraint from Ref.~\cite{Trott:2010iz} translates into the following approximate 2$\sigma$ bounds:
\begin{eqnarray}
	-1.1 \leq {\rm Re}(X_2 Y_2^*) \leq 0.7 \qquad {\rm for} \ M_{H_2^+} = 100 \ {\rm GeV}, \nonumber \\
	-4.0 \leq {\rm Re}(X_2 Y_2^*) \leq 2.6 \qquad {\rm for} \ M_{H_2^+} = 500 \ {\rm GeV}.
\end{eqnarray}	

Finally, Trott and Wise~\cite{Trott:2010iz}\footnote{The charged Higgs coupling parameters in Ref.~\cite{Trott:2010iz} are related to ours according to $\eta_U = -Y_2$, $\eta_D = -X_2^*$.} 
have considered the constraint on the CP-violating part of the charged Higgs couplings arising from the neutron electric dipole moment, using Naive Dimensional Analysis.  They obtain an upper bound of
\begin{equation}
	| {\rm Im} (X_2 Y_2^*) | \lesssim 0.1 \ (0.4) \qquad {\rm for} \ M_{H_2^+} = 100 \ (500) \ {\rm GeV}.
\end{equation}
While this is only an order-of-magnitude upper bound due to the use of the Naive Dimensional Analysis approximation, at face value it is somewhat stronger than the constraint on ${\rm Re}(X_2 Y_2^*)$.


\section{Couplings of the other charged Higgs boson}
\label{sec:H3}

We now turn to the couplings of $H_3^+$.  Starting from the three relations obtained from $\sum_i U^{\dagger}_{fi} U_{if^{\prime}} = \delta_{ff^{\prime}}$ with $f \neq f^{\prime}$, we can solve for the $H_3^+$ coupling factors $X_3$, $Y_3$, and $Z_3$ in terms of $X_2$, $Y_2$, and $Z_2$.  We use the phase freedom of $H_3^+$ to choose $Z_3$ to be real.

The magnitude and phase of $X_3$ are obtained from
\begin{equation}
	X_3^2 = (-1 - X_2 Z_2) \frac{(1 - X_2 Y_2^*)}{(1 - Y_2^* Z_2)}, \qquad 
	|X_3|^2 = (-1 - X_2^* Z_2) \frac{(1 - X_2 Y_2^*)}{(1 - Y_2^* Z_2)}.
	\label{eq:H3a}
\end{equation} 
The magnitude and phase of $Y_3$ are obtained from
\begin{equation}
	Y_3^2 = (1 - Y_2 Z_2) \frac{(1 - X_2^* Y_2)}{(-1 - X_2^* Z_2)}, \qquad
	|Y_3|^2 = (1 - Y_2^* Z_2) \frac{(1 - X_2^* Y_2)}{(-1 - X_2^* Z_2)}.
	\label{eq:H3b}
\end{equation}
The real parameter $Z_3$ is obtained from
\begin{equation}
	Z_3^2 = (1 - Y_2^* Z_2) \frac{(-1 - X_2 Z_2)}{(1 - X_2 Y_2^*)}.
	\label{eq:H3c}
\end{equation}
Note that the expressions for $|X_3|^2$, $|Y_3|^2$, and $Z_3^2$ are written in terms of the complex couplings $X_2$ and $Y_2$, yet in each case they yield the value of a real parameter.  The imaginary parts of the right-hand sides of these expressions must thus be zero, providing an alternate form of the constraint among the five real parameters in $X_2$, $Y_2$, and $Z_2$.

Convenient expressions for the vevs can also be obtained from the unitarity relation $\sum_i U^{\dagger}_{fi} U_{if} = 1$:
\begin{eqnarray}
	v_d^2 &=& \frac{v_{\rm SM}^2}{1 + |X_2|^2 + |X_3|^2} 
	= \frac{v_{\rm SM}^2}{1 + |X_2|^2 
	+ \left[ (-1 - X_2^* Z_2) (1 - X_2 Y_2^*)/(1 - Y_2^* Z_2) \right]},
	\nonumber \\
	v_u^2 &=& \frac{v_{\rm SM}^2}{1 + |Y_2|^2 + |Y_3|^2} 
	= \frac{v_{\rm SM}^2}{1 + |Y_2|^2 
	+ \left[ (1 - Y_2^* Z_2) (1 - X_2^* Y_2)/(-1 - X_2^* Z_2) \right]},
	\nonumber \\
	v_{\ell}^2 &=& \frac{v_{\rm SM}^2}{1 + |Z_2|^2 + |Z_3|^2} 
	= \frac{v_{\rm SM}^2}{1 + |Z_2|^2 
	+ \left[ (1 - Y_2^* Z_2) (-1 - X_2 Z_2)/(1 - X_2 Y_2^*) \right]},
\end{eqnarray}
where in the second expression for each vev we made use of Eqs.~(\ref{eq:H3a}), (\ref{eq:H3b}), and~(\ref{eq:H3c}).

\section{Adding a fourth Higgs doublet}
\label{sec:4HDM}

We have seen that the democratic 3HDM reproduces the charged Higgs coupling freedom of the Yukawa-aligned 2HDM but with two significant constraints: first, that the magnitudes of the coupling parameters cannot all be greater than one or all be less than one; and second, that the phase of the second complex coupling parameter is fixed in terms of the magnitudes of the three parameters and the phase of the first.  These constraints arise from the unitarity of the $3\times 3$ charged Higgs mixing matrix; the second one in particular comes from the fact that the $3\times 3$ mixing matrix is parameterized in terms of three angles and only one phase.

We now show that both these constraints are removed if we extend the Higgs sector by adding a fourth Higgs doublet $\Phi_0$, carrying a vev $v_0$, and with no couplings to fermions in accordance with natural flavor conservation.

In the presence of a fourth doublet, the $3\times 3$ charged Higgs mixing matrix $U$ of the democratic 3HDM becomes a $4\times 4$ matrix, which we denote $\widetilde U$:
\begin{equation}
	\left( \begin{array}{c} G^+ \\ H_2^+ \\ H_3^+ \\ H_4^+ \end{array} \right)
	= \widetilde U 
	\left( \begin{array}{c} \phi_d^+ \\ \phi_u^+ \\ \phi_\ell^+ \\ \phi_0^+ \end{array} \right).
\end{equation}
The unitarity constraint on the first row of $\widetilde U$ yields the usual sum rule for the vevs:
\begin{equation}
	|\widetilde U_{1d}|^2 + |\widetilde U_{1u}|^2 + |\widetilde U_{1\ell}|^2 
	+ |\widetilde U_{10}|^2 
	= \frac{v_d^2}{v_{\rm SM}^2} + \frac{v_u^2}{v_{\rm SM}^2} 
	+ \frac{v_{\ell}^2}{v_{\rm SM}^2} + \frac{v_0^2}{v_{\rm SM}^2}
	= 1.
\end{equation}
The unitarity constraint on the second row of $\widetilde U$ yields,
\begin{equation}
	|X_2|^2 |\widetilde U_{1d}|^2 + |Y_2|^2 |\widetilde U_{1u}|^2 
	+ |Z_2|^2 |\widetilde U_{1 \ell}|^2 + |\widetilde U_{20}|^2 = 1,
\end{equation}
where $X_2$, $Y_2$, and $Z_2$ are defined as in Eq.~(\ref{eq:xyz}) with $U$ replaced by $\widetilde U$.  For a given choice of vevs, this expression constrains $|X_2|^2$, $|Y_2|^2$, and $|Z_2|^2$ to lie in a {\it volume} in the three-dimensional positive-definite parameter space, extending from the origin (for $|\widetilde U_{20}|^2 = 1$) up to the plane obtained by setting $|\widetilde U_{20}|^2 = 0$; this plane passes through the point $(a,a,a)$ with $a = (1 - v_0^2/v_{\rm SM}^2)^{-1} \geq 1$.  Increasing $v_0$ from zero to $v_{\rm SM}$ moves this intersection point from $(1,1,1)$ out to infinity.  Therefore there is no theoretical constraint on the magnitudes of $X_2$, $Y_2$, and $Z_2$ in the democratic model with four doublets.

We now consider the complex phases.  After using the rephasing freedom of $\phi_f^+$ and $H_i^+$, the $4\times 4$ unitary matrix is parameterized by six angles and three phases; in particular, there are enough free phase parameters for the phases of $X_2$ and $Y_2$ to be independent free parameters.  We can check this through an explicit parameterization of $\widetilde U$.  Choosing the basis for $\phi_f^+$ in which the vevs are real and defining $\tan\beta = v_u/v_d$, $\tan \gamma = \sqrt{v_d^2 + v_u^2}/v_{\ell}$, and $\tan \omega = \sqrt{v_d^2 + v_u^2 + v_{\ell}^2}/v_0$, we have\footnote{We have not yet applied the phase rotation to $H_2^+$ needed to make $Z_2$ real.}
\begin{eqnarray}
	\widetilde U &=& \left( \begin{array}{cccc}
		1 & 0 & 0 & 0 \\
		0 & 1 & 0 & 0 \\
		0 & 0 & c_3 & s_3 e^{i\delta_3} \\
		0 & 0 & -s_3 e^{-i \delta_3} & c_3 \end{array} \right)
	\left( \begin{array}{cccc}
		1 & 0 & 0 & 0 \\
		0 & c_2 & 0 & s_2 e^{i \delta_2} \\
		0 & 0 & 1 & 0 \\
		0 & -s_2 e^{-i \delta_2} & 0 & c_2 \end{array} \right)
	\left( \begin{array}{cccc}
		1 & 0 & 0 & 0 \\
		0 & c_1 & s_1 e^{i \delta_1} & 0 \\
		0 & -s_1 e^{-i \delta_1} & c_1 & 0 \\
		0 & 0 & 0 & 1 \end{array} \right) \nonumber \\
	&& \times \left( \begin{array}{cccc}
		s_\omega & 0 & 0 & c_\omega \\
		0 & 1 & 0 & 0 \\
		0 & 0 & 1 & 0 \\
		-c_\omega & 0 & 0 & s_\omega \end{array} \right)
	\left( \begin{array}{cccc}
		s_\gamma & 0 & c_\gamma & 0 \\
		0 & 1 & 0 & 0 \\
		-c_\gamma & 0 & s_\gamma & 0 \\
		0 & 0 & 0 & 1 \end{array} \right)
	\left( \begin{array}{cccc}
		c_\beta & s_\beta & 0 & 0 \\
		-s_\beta & c_\beta & 0 & 0 \\
		0 & 0 & 1 & 0 \\
		0 & 0 & 0 & 1 \end{array} \right),
	\label{eq:Utilde}
\end{eqnarray}
where $s_i$, $c_i$ with $i = 1, 2, 3$ denote the sine and cosine of angles $\theta_{1,2,3}$, and $\delta_{1,2,3}$ are three complex phases.  Here the last three matrices determine the charged Goldstone boson while the first three accomplish the diagonalization of the remaining three physical charged Higgs mass eigenstates.  Evaluating Eq.~(\ref{eq:Utilde}), we obtain the elements of $\widetilde U$ that enter $X_2$, $Y_2$, and $Z_2$ (we have not yet applied the phase rotation to $H_2^+$ needed to make $\widetilde U_{2 \ell}$ real):
\begin{eqnarray}
	\widetilde U_{2d} &=& -c_1 c_2 s_\beta
	- (s_1 c_2 e^{i\delta_1} c_\gamma + s_2 e^{i\delta_2} c_\omega s_\gamma) c_\beta,
	\nonumber \\
	\widetilde U_{2u} &=& c_1 c_2 c_\beta 
	- (s_1 c_2 e^{i\delta_1} c_\gamma + s_2 e^{i\delta_2} c_\omega s_\gamma) s_\beta,
	\nonumber \\
	\widetilde U_{2\ell} &=& s_1 c_2 e^{i \delta_1} s_\gamma - s_2 e^{i \delta_2} c_\omega c_\gamma.
\end{eqnarray}
In particular, the parameter freedom is such that it is not possible to solve for one of the five real coupling degrees of freedom in terms of the other four.

Thus we see that the four-Higgs-doublet extension of the democratic 3HDM fully reproduces the charged Higgs coupling parameters of the Yukawa-aligned 2HDM without any theoretical constraints.

\section{Discussion and conclusions}
\label{sec:conclusions}

In this paper we studied the charged Higgs sector of the democratic 3HDM, in which one doublet couples to down-type quarks, one to up-type quarks, and one to charged leptons.  This model, in which flavor-changing neutral Higgs couplings are avoided by the imposition of natural flavor conservation, very nearly reproduces the Yukawa coupling structure of the charged Higgs in the Yukawa-aligned 2HDM, in which flavor-changing neutral Higgs couplings are avoided through a linear realization of minimal flavor violation (called Yukawa alignment).
We implemented a general parameterization of the couplings of the two physical charged Higgs bosons to fermions, and showed that the couplings of the lighter charged Higgs $H_2^+$ can be written in terms of two complex and one real parameters $X_2$, $Y_2$, and $Z_2$.  Unitarity constraints on the mixing matrix for the charged Higgs bosons require that one of these five real parameters (three magnitudes and two phases) is fixed in terms of the other four.  Unitarity also requires that the magnitudes of $X_2$, $Y_2$, and $Z_2$ are not all greater than one or all less than one.  
These two subtle constraints distinguish the couplings of $H_2^+$ in the democratic 3HDM from those in the Yukawa-aligned 2HDM, in which the three magnitudes and two phases of $X_2$, $Y_2$, and $Z_2$ are all theoretically unconstrained.  Adding a fourth Higgs doublet with no couplings to fermions removes these two constraints, reproducing the full coupling parameter freedom of the Yukawa-aligned 2HDM.

How else can we experimentally distinguish between the models?  Clearly, discovery of a second physical charged Higgs boson $H_3^+$, or of neutral Higgs bosons beyond the three predicted in a 2HDM, rules out the minimal Yukawa-aligned 2HDM.  A more subtle test involves the couplings of the neutral Higgs bosons.  The Yukawa-aligned 2HDM contains three neutral Higgs states, $S_{1,2,3}^0$, which are admixtures of the two CP-even and one physical CP-odd states of the two Higgs doublets~\cite{Pich:2009sp}.  The couplings of these three neutral Higgs bosons to fermions are fixed in terms of the SM Yukawa matrices, the same three coupling parameters $X_2$, $Y_2$, and $Z_2$ that appear in the charged Higgs sector, and the unitary matrix that diagonalizes the mass-squared matrix for $S_{1,2,3}^0$.  

A model with three or more Higgs doublets contains not only additional charged Higgs states, but two additional neutral Higgs degrees of freedom per doublet.  Consider a rotation of the doublets to the ``charged-Higgs basis'':
\begin{equation}
	H_1 = \left( \begin{array}{c} G^+ \\ 
	(v_{\rm SM} + \phi_1^{0,r} + i G^0)/\sqrt{2} \end{array} \right), \qquad
	H_2 = \left( \begin{array}{c} H_2^+ \\
	(\phi_2^{0,r} + i \phi_2^{0,i})/\sqrt{2} \end{array} \right), \qquad
	H_3 = \left( \begin{array}{c} H_3^+ \\
	(\phi_3^{0,r} + i \phi_3^{0,i})/\sqrt{2} \end{array} \right), \ \ \cdots
\end{equation}
If the first three discovered neutral Higgs states $S_{1,2,3}^0$ are mixtures of {\it only} $\phi_1^{0,r}$, $\phi_2^{0,r}$, and $\phi_2^{0,i}$, then their couplings to fermions will depend on the same set of parameters as the charged Higgs couplings, just as in the Yukawa-aligned 2HDM.  
However, if $S_{1,2,3}^0$ contain admixtures of the neutral Higgs states from $H_3$ (and/or other additional doublets), their couplings to fermions will depend on new additional parameters---in particular, the coupling factors $X_3$, $Y_3$, and $Z_3$ of the charged Higgs state $H_3^+$.  (In the special case of three doublets, $X_3$, $Y_3$, and $Z_3$ are constrained directly by the couplings $X_2$, $Y_2$, and $Z_2$ as shown in Sec.~\ref{sec:H3}.)

Furthermore, the SU(2)$_L$ gauge couplings of $S_{1,2,3}^0$ to $W^+W^-$ and to $W^+ H_2^-$ obey sum rules.  Writing the Feynman rules with all particles incoming as $i g_{S_i^0 W^- W^+} g_{\mu\nu}$ and $i g_{S_i^0 H_2^- W^+} (p_S - p_H)_{\mu}$, where $p_S$ and $p_H$ are the incoming momenta of $S_i^0$ and $H_2^-$, respectively, we have for the states in the charged-Higgs basis:
\begin{eqnarray}
	&& g_{\phi_1^{0,r} W^- W^+} = g M_W, \qquad
	g_{\phi_2^{0,r} W^- W^+} = 0, \qquad
	g_{\phi_2^{0,i} W^- W^+} = 0, \nonumber \\
	&& g_{\phi_1^{0,r} H_2^- W^+} = 0, \qquad
	g_{\phi_2^{0,r} H_2^- W^+} = -g/2, \qquad
	g_{\phi_2^{0,i} H_2^- W^+} = -ig/2.
\end{eqnarray}
If $S_{1,2,3}^0$ are mixtures of {\it only} $\phi_1^{0,r}$, $\phi_2^{0,r}$, and $\phi_2^{0,i}$, we obtain two sum rules for the squared magnitudes of these couplings:
\begin{equation}
	\sum_i |g_{S_i^0 W^+ W^-} |^2 = g^2 M_W^2, \qquad
	\sum_i |g_{S_i^0 H_2^- W^+} |^2 = g^2/2.
	\label{eq:sumrule}
\end{equation}
However, if $S_{1,2,3}^0$ contain admixtures of the neutral Higgs states from $H_3$ (and/or other additional doublets), one or both of the sum rules in Eq.~(\ref{eq:sumrule}) will not be saturated, indicating that $S_{1,2,3}^0$ do not together capture all of the states $\phi_1^{0,r}$, $\phi_2^{0,r}$, and $\phi_2^{0,i}$.

We have thus seen that distinguishing fundamental Yukawa alignment from Higgs-sector-based natural flavor conservation can be very challenging.  It will require a detailed study of charged Higgs couplings as well as a thorough exploration of the weak scale for additional Higgs doublets.


\begin{acknowledgments}
We thank Andrew Akeroyd, Martin Jung, and Michael Trott for helpful comments.
H.E.L.\ was partially supported by the Natural Sciences and Engineering
Research Council of Canada.  
\end{acknowledgments}


\appendix

\section{Scalar potential of the democratic three-Higgs-doublet model}

The most general SU(2)$_L \times$U(1)$_Y$-invariant potential for three Higgs doublets, subject to the $Z_2$ symmetries in Table~\ref{tab:Z2charges}, can be written as
\begin{eqnarray}
	V &=& m_{uu}^2 \Phi_u^{\dagger} \Phi_u + m_{dd}^2 \Phi_d^{\dagger} \Phi_d 
	+ m_{\ell\ell}^2 \Phi_{\ell}^{\dagger} \Phi_{\ell}
	 - \left[ m_{ud}^2 \Phi_u^{\dagger} \Phi_d + m_{u\ell}^2 \Phi_u^{\dagger} \Phi_{\ell} 
	 + m_{d\ell}^2 \Phi_d^{\dagger} \Phi_{\ell} + {\rm h.c.} \right] 
	 \nonumber \\
	 && + \frac{1}{2} \lambda_u (\Phi_u^{\dagger} \Phi_u)^2
	 + \frac{1}{2} \lambda_d (\Phi_d^{\dagger} \Phi_d)^2
	 + \frac{1}{2} \lambda_{\ell} (\Phi_{\ell}^{\dagger} \Phi_{\ell})^2
	 \nonumber \\
	 && + \lambda_{ud} (\Phi_u^{\dagger} \Phi_u)(\Phi_d^{\dagger} \Phi_d)
	 + \lambda_{u\ell} (\Phi_u^{\dagger} \Phi_u)(\Phi_{\ell}^{\dagger} \Phi_{\ell})
	 + \lambda_{d\ell} (\Phi_d^{\dagger} \Phi_d)(\Phi_{\ell}^{\dagger} \Phi_{\ell})
	 \nonumber \\
	 && + \lambda_{ud}^{\prime} (\Phi_u^{\dagger} \Phi_d)(\Phi_d^{\dagger} \Phi_u)
	 + \lambda_{u\ell}^{\prime} (\Phi_u^{\dagger} \Phi_{\ell})(\Phi_{\ell}^{\dagger} \Phi_u)
	 + \lambda_{d\ell}^{\prime} (\Phi_d^{\dagger} \Phi_{\ell})(\Phi_{\ell}^{\dagger} \Phi_d)
	 \nonumber \\
	 && + \frac{1}{2} \left[ 
	 \lambda_{ud}^{\prime \prime} (\Phi_u^{\dagger} \Phi_d)^2
	 + \lambda_{u\ell}^{\prime \prime} (\Phi_u^{\dagger} \Phi_{\ell})^2
	 + \lambda_{d\ell}^{\prime \prime} (\Phi_d^{\dagger} \Phi_{\ell})^2
	 + {\rm h.c.} \right],
\end{eqnarray}
where we have retained the terms proportional to $m_{ud}^2$, $m_{u\ell}^2$, and $m_{d\ell}^2$ that break the $Z_2$ symmetries softly.\footnote{We note that omitting these soft-$Z_2$-breaking terms does not change our conclusions about the complex structure of the charged Higgs mixing matrix; we will keep them here for generality and because they allow the model to have a decoupling limit in which the extra Higgs states are taken heavy without requiring large quartic couplings.}  This potential contains six complex parameters: the soft-$Z_2$-breaking mass-squared terms $m_{ud}^2$, $m_{u\ell}^2$, and $m_{d\ell}^2$, and the quartic couplings $\lambda_{ud}^{\prime\prime}$, $\lambda_{u\ell}^{\prime\prime}$, and $\lambda_{d\ell}^{\prime\prime}$.

This potential is invariant under a common global phase rotation of $\Phi_u$, $\Phi_d$, and $\Phi_{\ell}$ (as required by U(1)$_Y$-invariance); we can use this to choose $v_{\ell}$ real and positive without any loss of generality.  Performing a phase rotation on $\Phi_u$ and $\Phi_d$ to make $v_u$ and $v_d$ real and positive imposes two relations among the imaginary parts of the complex parameters of the potential:
\begin{eqnarray}
	{\rm Im}(m_{u\ell}^2) &=& - \frac{v_d}{v_{\ell}} {\rm Im}(m_{ud}^2) 
	+ \frac{v_u v_d^2}{2 v_{\ell}} {\rm Im}(\lambda_{ud}^{\prime\prime})
	+ \frac{v_u v_{\ell}}{2} {\rm Im}(\lambda_{u\ell}^{\prime\prime}),
	\nonumber \\ 
	{\rm Im}(m_{d\ell}^2) &=& \frac{v_u}{v_{\ell}} {\rm Im}(m_{ud}^2) 
	- \frac{v_u^2 v_d}{2 v_{\ell}} {\rm Im}(\lambda_{ud}^{\prime\prime})
	+ \frac{v_d v_{\ell}}{2} {\rm Im}(\lambda_{d\ell}^{\prime\prime}).
\end{eqnarray}
Minimizing the potential allows three more parameters to be eliminated in favor of the vevs:
\begin{eqnarray}
	m_{uu}^2 &=& \frac{v_d}{v_u} {\rm Re}(m_{ud}^2) 
	+ \frac{v_{\ell}}{v_u} {\rm Re}(m_{u\ell}^2)
	- \frac{v_u^2}{2} \lambda_u
	- \frac{v_d^2}{2} \left[ \lambda_{ud} + \lambda_{ud}^{\prime} + {\rm Re}(\lambda_{ud}^{\prime\prime}) \right] 
	- \frac{v_{\ell}^2}{2} \left[ \lambda_{u\ell} + \lambda_{u\ell}^{\prime} + {\rm Re}(\lambda_{u\ell}^{\prime\prime}) \right],
	\nonumber \\
	m_{dd}^2 &=& \frac{v_u}{v_d} {\rm Re}(m_{ud}^2) 
	+ \frac{v_{\ell}}{v_d} {\rm Re}(m_{d\ell}^2)
	- \frac{v_d^2}{2} \lambda_d
	- \frac{v_u^2}{2} \left[ \lambda_{ud} + \lambda_{ud}^{\prime} + {\rm Re}(\lambda_{ud}^{\prime\prime}) \right]
	- \frac{v_{\ell}^2}{2} \left[ \lambda_{d\ell} + \lambda_{d\ell}^{\prime} + {\rm Re}(\lambda_{d\ell}^{\prime\prime}) \right],
	\nonumber \\
	m_{\ell\ell}^2 &=& \frac{v_u}{v_{\ell}} {\rm Re}(m_{u\ell}^2)
	+ \frac{v_d}{v_{\ell}} {\rm Re}(m_{d\ell}^2)
	- \frac{v_{\ell}^2}{2} \lambda_{\ell}
	- \frac{v_u^2}{2} \left[ \lambda_{u\ell} + \lambda_{u\ell}^{\prime} + {\rm Re}(\lambda_{u\ell}^{\prime\prime}) \right]
	- \frac{v_d^2}{2} \left[ \lambda_{d\ell} + \lambda_{d\ell}^{\prime} + {\rm Re}(\lambda_{d\ell}^{\prime\prime}) \right].
\end{eqnarray}

Applying these conditions we find the terms in the potential that are bilinear in the charged scalar fields:
\begin{eqnarray}
	V &\supset& \phi_u^- \phi_u^+ \left[ \frac{v_d}{v_u} A_{ud} + \frac{v_{\ell}}{v_u} A_{u\ell} \right]
	+ \phi_d^- \phi_d^+ \left[ \frac{v_u}{v_d} A_{ud} + \frac{v_{\ell}}{v_d} A_{d\ell} \right]
	+ \phi_{\ell}^- \phi_{\ell}^+ \left[ \frac{v_u}{v_{\ell}} A_{u\ell} + \frac{v_d}{v_{\ell}} A_{d\ell} \right]
	\nonumber \\
	&& + \left\{ \phi_u^- \phi_d^+ \left[ -A_{ud} - iB \right] 
	+ \phi_u^- \phi_{\ell}^+ \left[ -A_{u\ell} + i \frac{v_d}{v_{\ell}} B \right] 
	+ \phi_d^- \phi_{\ell}^+ \left[ -A_{d\ell} - i \frac{v_u}{v_{\ell}} B \right] + {\rm h.c.} \right\},
\end{eqnarray}
where
\begin{eqnarray}
	A_{ud} &=& {\rm Re}(m_{ud}^2) - \frac{v_u v_d}{2} \left[ \lambda_{ud}^{\prime} + {\rm Re}(\lambda_{ud}^{\prime\prime}) \right], 
	\nonumber \\
	A_{u\ell} &=& {\rm Re}(m_{u\ell}^2) - \frac{v_u v_{\ell}}{2} \left[ \lambda_{u\ell}^{\prime} + {\rm Re}(\lambda_{u\ell}^{\prime\prime}) \right],
	\nonumber \\
	A_{d\ell} &=& {\rm Re}(m_{d\ell}^2) - \frac{v_d v_{\ell}}{2} \left[ \lambda_{d\ell}^{\prime} + {\rm Re}(\lambda_{d\ell}^{\prime\prime}) \right],
	\nonumber \\
	B &=& {\rm Im}(m_{ud}^2) - \frac{v_u v_d}{2} {\rm Im}(\lambda_{ud}^{\prime\prime}).
\end{eqnarray}

We diagonalize the resulting charged Higgs mass-squared matrix $\mathcal{M}^2$ in two stages by dividing the charged Higgs mixing matrix in Eq.~(\ref{eq:Udef}) according to $U = U_2 U_1$, with [see also Eq.~(\ref{eq:Uexplicit})]
\begin{equation}
	U_1 = \left( \begin{array}{ccc}
		s_\gamma & 0 & c_\gamma \\
		0 & 1 & 0 \\
		-c_\gamma & 0 & s_\gamma \end{array} \right)
		\left( \begin{array}{ccc}
		c_\beta & s_\beta & 0 \\
		-s_\beta & c_\beta & 0 \\
		0 & 0 & 1 \end{array} \right), \qquad \qquad
	U_2 = \left( \begin{array}{ccc} 
		1 & 0 & 0 \\
		0 & e^{-i \delta} & 0 \\
		0 & 0 & 1 \end{array} \right)
		\left( \begin{array}{ccc}
		1 & 0 & 0 \\
		0 & c_\theta & s_\theta e^{i \delta} \\
		0 & -s_\theta e^{-i \delta} & c_\theta \end{array} \right).
\end{equation}
The rotation $U_1$ isolates the charged Goldstone boson, yielding
\begin{equation}
	\mathcal{M}^{2 \prime} \equiv U_1 \mathcal{M}^2 U_1^{\dagger} 
	= \left( \begin{array}{ccc} 0 & 0 & 0 \\
	0 & \mathcal{M}^2_{22} & \mathcal{M}^2_{23} \\
	0 & \mathcal{M}^{2*}_{23} & \mathcal{M}^2_{33} \end{array} \right),
	\label{eq:Mprime}
\end{equation}
where 
\begin{eqnarray}
	\mathcal{M}^2_{22} &=& \frac{v_{ud}^2}{v_u v_d} A_{ud} 
	+ \frac{v_u^2 v_{\ell}}{v_d v_{ud}} A_{d\ell} 
	+ \frac{v_d^2 v_{\ell}}{v_u v_{ud}} A_{u\ell},
	\nonumber \\
	\mathcal{M}^2_{33} &=& \frac{v_d v_{\rm SM}^2}{v_{\ell} v_{ud}^2} A_{d\ell}
	+ \frac{v_u v_{\rm SM}^2}{v_{\ell} v_{ud}^2} A_{u\ell},
	\nonumber \\
	\mathcal{M}^2_{23} &=& \frac{v_u v_{\rm SM}}{v_{ud}^2} A_{d\ell}
	- \frac{v_d v_{\rm SM}}{v_{ud}^2} A_{u\ell}
	+ i \frac{v_{\rm SM}}{v_{\ell}} B,
\end{eqnarray}
and $v_{ud} \equiv \sqrt{v_u^2 + v_d^2}$.  Note that the (23) and (32) elements of this matrix are complex.  

Finally we determine the mixing angle $\theta$ and the phase $\delta$ by requiring that the rotation $U_2$ diagonalize the matrix in Eq.~(\ref{eq:Mprime}).  The phase is given by
\begin{equation}
	\delta = {\rm phase}(\mathcal{M}^2_{23}),
\end{equation}
with $0 \leq \delta < 2\pi$.  Choosing $H_2^+$ to be lighter than $H_3^+$ yields the mass eigenstates,
\begin{equation}
	M^2_{H_2^+, H_3^+} = \frac{1}{2} \left[ \mathcal{M}_{22}^2 + \mathcal{M}_{33}^2
	\mp \sqrt{ (\mathcal{M}_{22}^2 - \mathcal{M}_{33}^2)^2 + 4 | \mathcal{M}_{23}^2 |^2} \right],
\end{equation}
and the mixing angle $\theta$,
\begin{equation}
	\sin 2\theta = \frac{ -2 |\mathcal{M}_{23}^2|}
	{\sqrt{(\mathcal{M}_{22}^2 - \mathcal{M}_{33}^2)^2 + 4 | \mathcal{M}_{23}^2 |^2}},
	\qquad \qquad
	\cos 2 \theta = \frac{ -\mathcal{M}_{22}^2 + \mathcal{M}_{33}^2}
	{\sqrt{(\mathcal{M}_{22}^2 - \mathcal{M}_{33}^2)^2 + 4 | \mathcal{M}_{23}^2 |^2}},
\end{equation}
with $-\pi/2 \leq \theta \leq 0$.


\end{document}